\documentclass{PoS}

\title{SKA Deep Polarization and Cosmic Magnetism}

\ShortTitle{SKA Deep Polarization}

\author{\speaker{A.R. Taylor$^{a,b}$ },
Ivan Agudo$^c$,
Takuya Akahori$^d$,
Rainer Beck$^e$,
Bryan Gaensler$^f$,
George Heald$^g$,
Melanie Johnston-Hollitt$^h$,
Mathieu Langer$^i$,
Lawrence Rudnick$^j$,
Dongsu Ryu$^k$,
Anna Scaife$^l$,
Dominik Schleicher$^m$,
Jeroen Stil$^n$ \\
$^a$ University of Cape Town, E-mail: \email{russ@ast.uct.ac.ca} \\
$^b$ University of the Western Cape, E-mail: \email{russ@ast.uct.ac.ca} \\
$^c$ Joint Institute for VLBI in Europe, E-mail: \email{agudo@jive.nl} \\
$^d$ University of Sydney, E-mail: \email{akahori@physics.usyd.edu.au} \\
 $^e$ Max-Planck-Institut f\"ur Radioastronomy, E-mail: \email{rbeck@mpifr-bonn.mpg.de}\\  
 $^f$ University of Sydney, E-mail: \email{bryan.gaensler@sydney.edu.au}\\  
 $^g$ ASTRON, E-mail: \email{heald@astron.nl}\\  
 $^h$ Victoria University of Wellington, E-mail: \email{Melanie.Johnstron-Hollitt@vuw.ac.nz}\\  
 $^i$ Universit\'e Paris-Sud, E-mail: \email{mathieu.langer@ias.u-psud.fr}\\  
 $^j$ University of Minnesota E-mail: \email{larry@unm.edu}\\  
 $^k$ UNIST, E-mail: \email{ryu@sirius.unist.ac.kr}\\  
 $^l$University of Southampton, E-mail: \email{A.scaife@soton.ac.uk}\\  
 $^m$ Institut f\"ur Astrophysik G\"ottingen, E-mail: \email{dschleic@astro.physik.uni-goettingen.de}\\  
 $^n$ University of Calgary, E-mail: \email{jstil@ucalgary.ca}\\  
           }

\abstract{
Deep surveys with the SKA1-MID array offer for the first time the opportunity to systematically explore the polarization properties of 
the microJy source population. Our knowledge of the polarized sky approaching these levels is still very limited. In total intensity the 
population will be dominated by star-forming and normal galaxies to intermediate redshifts ($z \sim1-2$), and low-luminosity AGN to high redshift. 
The polarized emission from these objects is a powerful probe of their intrinsic magnetic fields and of their magnetic environments. 
For redshift of order 1 and above the broad bandwidth of the mid-bands span the Faraday thick and thin regimes allowing study of the 
intrinsic polarization properties of these objects as well as depolarization from embedded and foreground plasmas. The deep field 
polarization images will provide Rotation Measures data with very high solid angle density allowing a sensitive statistical analysis 
of the angular variation of RM on critical arc-minute scales from a magnetic component of Large Scale Structure of the Universe.}

\FullConference{
Advancing Astrophysics with the Square Kilometre Array\\
June 8-13, 2014\\
Giardini Naxos, Italy}

\newcommand{\skipthis}[1]{}

\begin{document}
\section{Introduction}

The implementation of broad bandwidths and new correlator capacities on the Jansky Very Large Array (JVLA) and 
the Giant Meterwave Radio Telescope (GMRT) are opening up polarization imaging 
of the radio sky at mid-frequencies to $\mu$Jy sensitivities (Rudnick \& Owen 2014, Taylor et al.\ 2014). 
However, the survey speeds of these instruments limit the amount of 
sky that can be imaged at these sensitivities in a reasonable time to a square degree or less, 
thus current studies of the polarized radio source population at these flux densities suffer from low number statistics. 
The MeerKAT SKA precursor continuum surveys will extend imaging at these sensitivities to larger 
areas (35 sq deg. to rms of 1.5\,$\mu$Jy), albeit with lower angular resolution. At 75 nJy rms and arc second resolution SKA1-MID can
take a step in sensitivity of over an order magnitude compared to the present state-of-the-art and planned precursor projects, and 
will survey a large enough area of sky to provide a statistical overview of the magnetic properties of source populations over cosmic
time.

\begin{figure}[bp]
\centering{
\includegraphics[width=1.04\textwidth]{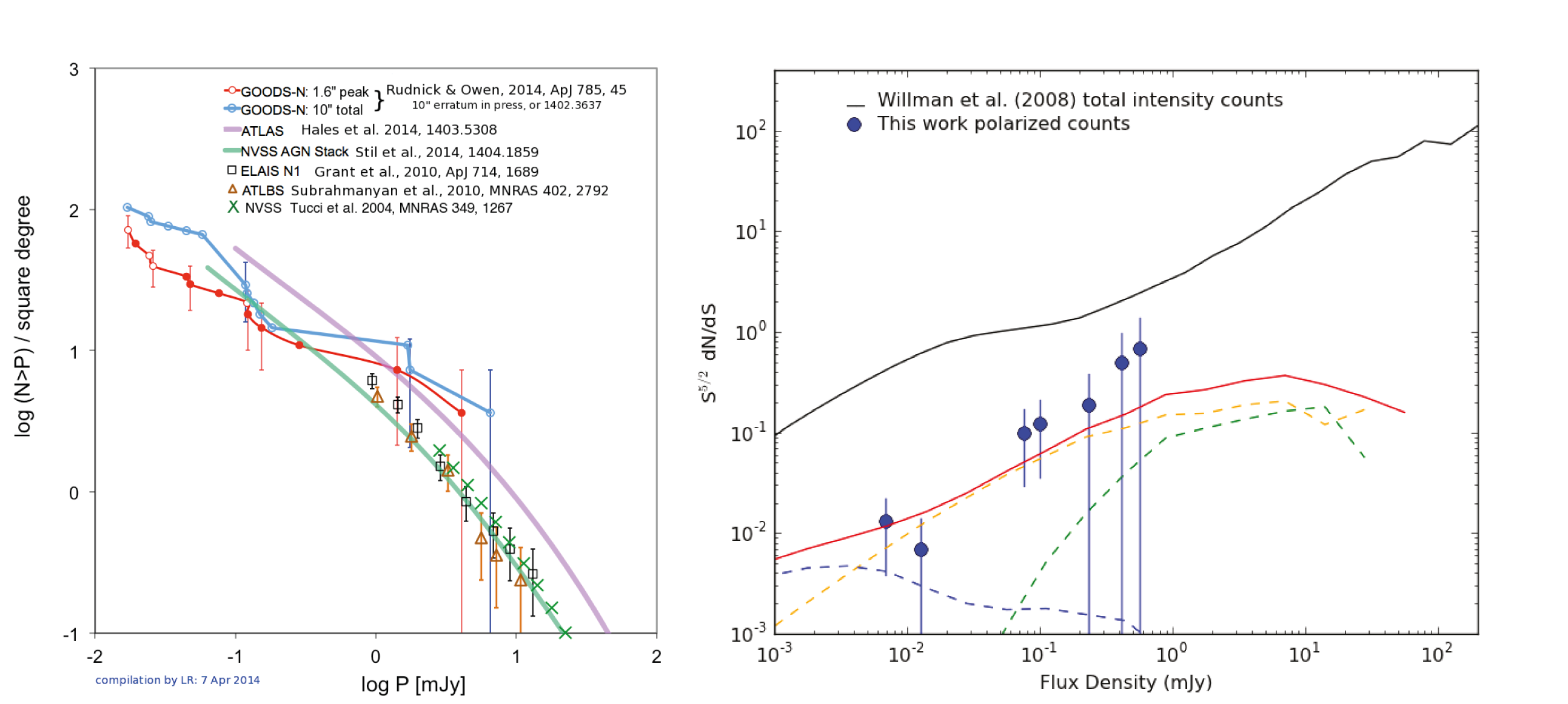}
\caption{Left: Compilation of cumulative polarized source counts down at 1.4 GHz to $\sim15\,\mu$Jy
from recent studies. Right: Differential polarized source counts at 5 GHz down to
5\,$\mu$Jy compared to predicted polarized counts of AGN (green and orange dashed lines) and 
disk galaxies (blue dashed line) \cite{Taylor14}.}
}
\label{taylor_fig1}
\end{figure}

Current studies hint at what the SKA1 deep polarized sky will reveal.  
Figure 1 shows at left a compilation of cumulative number counts of polarised sources at 1.4 GHz down to 15\,$\mu$Jy from recent
deep imaging projects. 
The plot demonstrates significant variation of results from different observers, likely due to both low number statistics and 
sensitivity to angular resolution. A slight flattening of the cumulative counts is suggested (Rudnick \& Owen 2014). 
This may reflect either a smaller overall number of polarized sources, perhaps arising from a decreasing fraction of 
Active Galactic Nuclei (AGN),
and/or an on-average smaller fractional polarization for fainter sources. 
The latter is consistent with integrated polarization properties of disk galaxies in which 
internal depolarization effects from thermal plasma decreases the fractional polarization at frequencies below a 
few GHz (Stil et al.\ 2009, Heald et al.\ 2014). Faint polarized counts are thus expected to be frequency dependent and broad-band data are
required to unravel the underlying astrophysics.
The right hand of Figure~\ref{taylor_fig1} shows polarized 
source counts at 5 GHz down to 5\,$\mu$Jy (Taylor et al.\  2014). The rising blue dashed line shows the predicted polarized counts for disk 
galaxies based on the polarization properties of nearby galaxies (Stil et al.\ 2009).  Deep imaging with SKA1 at GHz frequencies 
will probe the polarized properties of galaxies to high redshift.

\section{Emergence and Evolution of Magnetic Fields in Galaxies}

Through its ability to detect polarization of sources at high redshift, a deep survey is the cornerstone for investigation of the evolution of 
cosmic magnetic fields as it allows us to compare similar objects over the largest possible range in redshift, and place
observations of local samples (e.g. Beck et al.\ 2014) in an evolutionary context.
For disk galaxies, this range is z $\gtrsim$ 2.5, and for AGN and starbursts it reaches out to z $\gtrsim$ 7, 
into the epoch of reionization. 
Over this  period, galaxies formed and evolved, converting most of their gas into stars. 
The evolution of plasma and magnetic fields in 
galaxies is therefore closely related to the evolution of the cosmic star formation rate, and tied to the intergalactic medium through
accretion, galactic winds, tidal and ram stripping, and AGN activity.

With the SKA1 we can study magnetic fields over cosmic time through polarization of synchrotron emission in the source, 
Faraday rotation and depolarization, and Zeeman splitting. 
These typically require much higher sensitivity than detection of the source in total intensity. In 
the case of Faraday rotation and Zeeman splitting, chance alignment with a source in the background will allow detection of magnetic 
fields in sources that cannot be detected by their own emission. 
Since the background source must be at higher redshift, a deep field is more suitable to detect polarized
background sources for high-redshift objects. 

A majority of radio sources fainter than 100\,$\mu$Jy will be star-forming galaxies. The most luminous of these 
will be Ultra-luminous Infrared Galaxies (ULIRGs) and galaxy mergers. 
More common are relatively quiescent disk galaxies with a star formation 
rate that declines gradually by an order of magnitude from redshift 2 to the present. Deep surveys at other 
wavelengths are important to classify objects, and to measure the plasma density necessary to derive the magnetic
 field strength from Faraday rotation. In turn, a deep radio survey  is required to detect polarization in high-redshift 
 galaxies detected in infrared and optical surveys.

\begin{figure}[tbp]
\centering{
\includegraphics[width=1.05\textwidth]{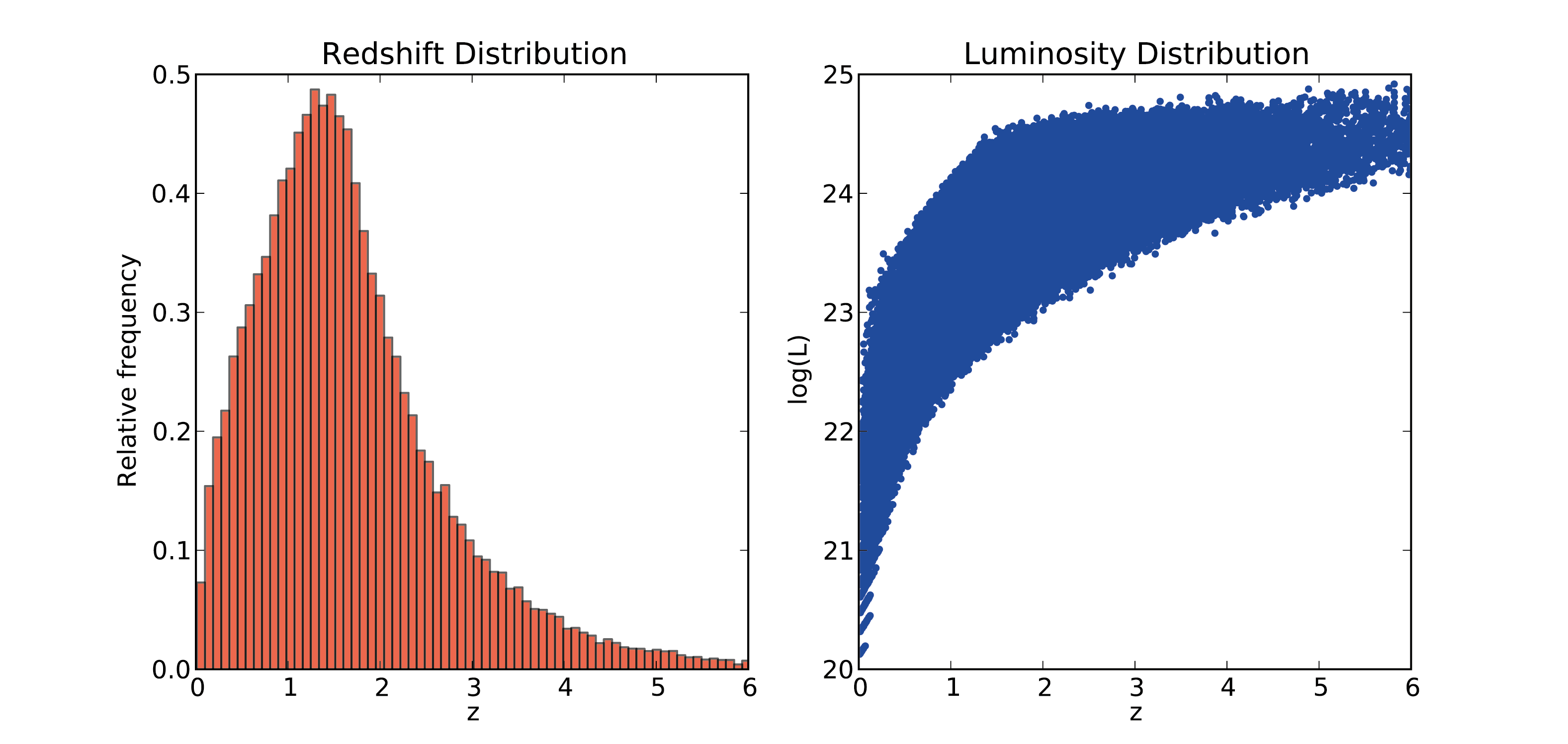}
\caption{Left: Redshift distribution of normal star-forming galaxies with polarization properties
of local disk galaxies that will have polarized flux density
greater than $10\sigma$ in the proposed SKA1 deep polarization field.  Right: redshift
versus radio luminosity of the same galaxies, showing that the galaxies probed are in the
bright tail of the radio luminosity function and have significant luminosity evolution.}
}
\label{taylor_fig2}
\end{figure}

For star forming galaxies, evolution of the magnetic field is closely related to evolution of the galaxy itself.
While $\mu$G strength 
magnetic fields with kpc scales should have formed in galaxy disks by z = 3, field ordering on the scale of a galaxy may have taken until 
z = 0.5, depending on galaxy mass (Arshakian et al.\ 2009).
Competing with this, galaxy interactions and continuous feedback by supernovae and stellar winds that 
enhance the turbulent component of the magnetic field, and may drive outflows that transport plasma and magnetic field 
from the disk into the halo. Since these processes scale with the global star formation rate, significant evolution is expected 
between z = 2 and the present. Also, the density of Faraday rotating plasma will gradually decrease over time as galaxies 
transform a significant fraction of their gaseous mass into stars, implying a gradual evolution in Faraday depth.
Figure~\ref{taylor_fig2} shows the predicted redshift distribution of galaxies with polarized flux density more than 10$\sigma$ 
in an SKA1-MID  band 2 \& 3 survey with sensitivity 75 nJy.  
These simulations are  based on models by Stil et al.\ (2009) for spiral galaxies at redshift 0, applied to normal star forming 
galaxies in the SKADS S3 simulation of Wilman et al.\ (2008).  
We expect to detect $\sim$5000 galaxies per square degree above 10$\sigma$.  A 10 square degree survey would thus provide
probes of the presence of ordered magnetic fields in of order 50,000 galaxies out to redshift > 4.
The luminosity- redshift diagram for these galaxies shows that many will have experienced significant 
luminosity evolution to redshift 0.  SKA1-MID is well-positioned to distinguish polarization 
properties of the more luminous regular star forming galaxies from those of their local (present day) counterparts. Zeeman 
splitting of HI absorption and Faraday rotation by systems in the line of sight to distant AGN will allow us to investigate 
magnetic fields in galaxies with radio emission below the detection limit of this survey.

Extreme starbursts in ULIRGS are environments that are very rare in the local universe. Little is known about magnetic fields 
in these objects. The dense interstellar medium suggests that any polarized emission related to star formation or an embedded 
AGN will be subject to very strong Faraday rotation. These objects may also have extended synchrotron halos associated with 
a galactic wind, or extended magnetic fields associated with tidal tails. These magnetic fields exist at the boundary of galactic 
and intergalactic space and may serve as seed fields for a more wide-spread intergalactic magnetic field.

\section{Magnetic Properties of AGN over Cosmic Time}
Active galactic nuclei are powered by supermassive black holes that accrete matter and are 
capable of forming pairs of relativistic jets that can extend up to distances far larger than the size of the host galaxy.  
The fact that jets in AGN are formed from magnetic systems, and that the magnetic field participating in their formation is 
partially dragged all along the jets, make polarization observations of their synchrotron emission a powerful tool to test 
magneto-hydrodynamic and emission models.
However, polarization studies of the radio AGN phenomenon have been limited by the lack of polarization purity, sensitivity, 
bandwidth and spectral resolution to make adequate studies of magnetic fields in AGN jets and their environments 
for massive samples of sources.
A deep full-polarimetric SKA1-MID survey on Bands 2 and 3 reaching a sensitivity of 75\,nJy/beam will have the 
ability to attack the long standing problem of the composition of AGN jets (i.e. the electron-proton content), and their plasma 
acceleration from large samples suited for statistical studies for the first time, see Agudo et al.\ (2014).
For that, it is essential to have available a circular polarization purity at least $\sim0.01$\,\%, since AGN typically show circular 
polarization moduli in the range $0.1 -1$\,\% at centimetre radio wavelengths.

The first large scale investigations on the polarization of AGN sources have been based on the 
NRAO VLA Sky Survey (NVSS) (Condon et al.\ 1998), 
which has detected 1.8 million extragalactic sources in total intensity. 
From these, $14\%$ show a clear polarization signal of at least $3\sigma$, and the majority of polarized 
extragalactic sources in the NVSS are AGN. 
Bright sources above $100$~mJy correspond to the radio-loud FRII sources (Fanaroff \& Riley 1974), and a 
gradual transition towards FRI sources occurs around $30$~mJy. 
Around a flux density of $1$~mJy, the contribution from star-forming galaxies becomes more relevant. 
So far, typical deep imaging projects have been restricted to flux density sensitivities of $\sim10\mu$J in total intensity 
(Hopkins et al. 2003). 
While no dependence of polarization properties on redshifts out to $z\sim3$ has been found, it is very interesting that  
surveys, including the NVSS, and the Dominion Radio Astrophysical Observatory Deep ELAIS N1 Field 
indicate an increase in fractional polarization with decreasing flux density (Mesa et al.\ 2002, Taylor et al.\ 2007). 
This anti-correlation suggests either a) a change in the magnetic field structure of the observed sources or b) a change of the 
properties their Faraday screen, perhaps related to the close intergalactic environment of the sources. 
Different solutions for this anti-correlation have been suggested, reflecting either a change in the population (Mesa et al.\ 2002), 
a change in the fraction of radio-quiet AGN or the FRII-FRI transition. 
From detailed studies of samples with high polarization, no strong dependence on optical morphology, redshift, linear size or 
radio power has been found (Shi et al.\ 2010), while a stacking analysis confirmed that the observed correlation continues
toward very low flux densities in the NVSS (Stil et al.\ 2014).
SKA1-MID will provide the means to take a major next step in this research, i.e. a deep polarimetric survey of the AGN 
populations up to the highest achievable sensitivities.

The superb sensitivity of SKA1 opens a new window to study the AGN phenomenon and its surrounding 
medium, from a cosmological perspective, back to the epoch of reionization, including the 
potential appearance of radio galaxies near or above $z\sim10$, at a time where the densities in the Universe were 
considerably  enhanced by a factor of $\sim10^3$. 
Due to the enhanced densities in the interstellar and intergalactic medium, it becomes more difficult for jets to break out from 
their host galaxy (Falcke et al.\ 2004). 
Such systems known as GHz-Peaked-Spectrum (GPS) sources have been observed in the local Universe, corresponding to 
environments that are either very dense or where the jet is very young (ODea 1998). 
Due to the characteristic evolution of angular scales and observed frequency with redshift, the population of high-z GPS sources 
will likely be shifted towards lower frequencies, but will show sizes compared to those of their local counterparts (Falcke et al.\ 2004).
The latter peak their radio spectrum at $\sim1$\,GHz, hence making a deep SKA1-MID total intensity survey between 
1 and 3\,GHz less suited for searches of high-z GPS than a parallel SKA1-LOW deep survey.
In contrast, a deep and wide band survey at frequencies $>1$\,GHz, where the sources are expected to be a 
factor of $\sim10$  fainter than in the SKA1-LOW bands, avoids strong Faraday depolarization by the dense environments
 in which the first AGN are expected to be embedded, and therefore perhaps allowing the detection of their linear polarization.  
A deep-polarimetric and wide-band SKA1-MID survey in Bands 2 and 3 hence opens the exciting 
possibility not only to probe the magnetic fields in the first AGN jets, but also their immediate dense intergalactic medium at $z\sim10$ .
 
\section{Detecting Magnetic Fields in the Cosmic Web}

 The SKA1 deep polarization survey will provide the supremely-dense RM grid of polarization probes of the intergalactic medium 
 required to detect and measure the properties of magnetic fields embedded in the large scale structure of the universe.
 The LCDM cosmology predicts the cosmic web of galaxy clusters and filaments. While clusters contain hot plasma of 
 T $>10^7$ K,  filaments are filled with plasma of $10^7$ K $>$ T $>10^5$ K which is referred to as the Warm Hot 
 Intergalactic Medium (WHIM). 
 The  plasmas are expected to be magnetized; diverse processes for seed magnetic fields have been suggested, and the seed 
 fields can be further amplified through compression and turbulent dynamo as well as leakage of galactic media during the 
 hierarchical structure formation in the Universe. See Ryu et al.\ (2012) and Widraw et al.\ (2012) for reviews.
 
The Intergalactic Magnetic Field (IGMF) in the cosmic web plays important roles in various astrophysical phenomena. 
For instance, the IGMF imprints its own existence on the cosmic microwave background, deflects the trajectory of cosmic-rays through 
the cosmic web, and it affects the
 thermal and dynamical evolutions of galaxy clusters. In addition, the IGMF may have provided seed fields of galaxies and 
 influenced the origin and nature of magnetic fields in spiral galaxies. 
  
\begin{figure}[tbp]
\centering{
\includegraphics[width=0.95\textwidth]{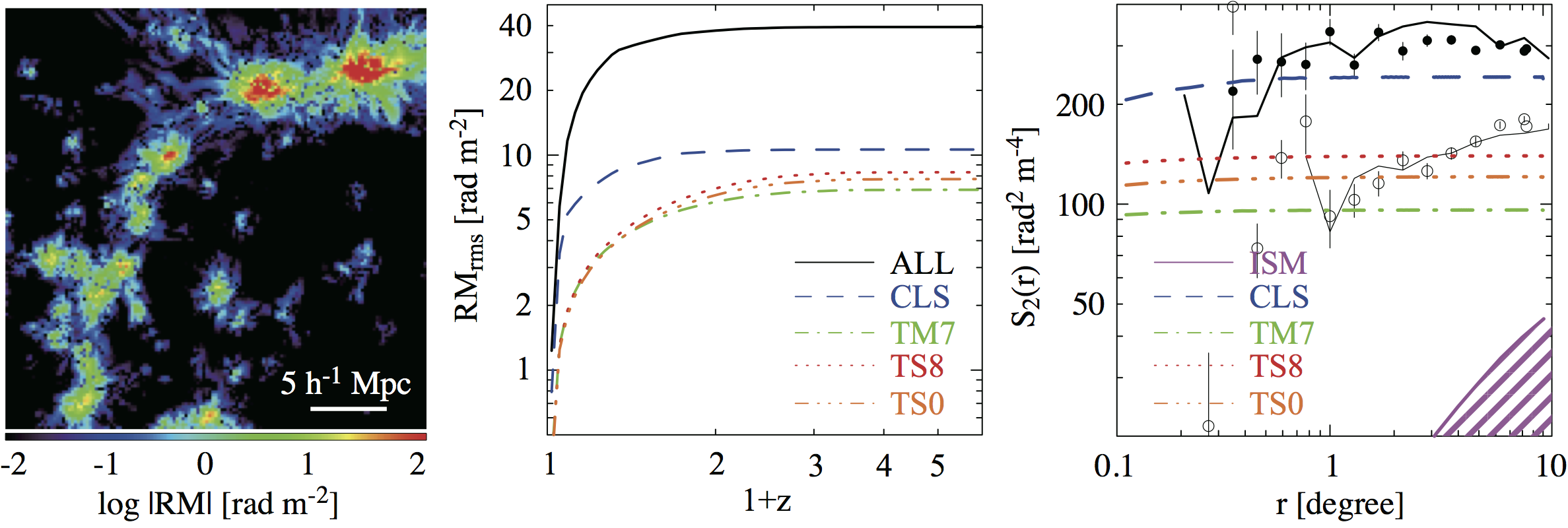}
\caption{Left: The RM map of the local universe of a $100$ $h^{-1}$Mpc depth (Akahori \& Ryu 2010). 
Middle: The rms value of RM integrated up to $z = 5$ (Akahori \& Ryu 2011). 
Results with different methods for cluster subtractions are shown as different colors. 
Right: Second-order structure functions of RM. Black circles (Mao et al.\ 2010) 
and lines (Stil.\ et al. 2011) are observed ones for $\sim 900$ deg$^2$ fields-of-view toward 
North (filled, thick) and South (open, thin) Galactic poles. The purple area indicates possible 
amplitudes for the Milky Way RM toward the poles (Akahori et al.\ 2013).} \label{taylor_fig3}
}

\end{figure}

RM has been the main tool for studies of the IGMF. It has revealed $B \sim 1-10$\, $\mu$G in clusters.
Detection of magnetic fields in filaments is challenged due to the expected very low value of RM. 
Cosmological simulations suggest the IGMF  in filaments may have $B \sim 1 - 100$ nG (Ryu et al.\ 2008). 
 With the IGMF, the rms value of RM, RM$_{\rm rms}$, through a single filament would be $\sim$1 rad\,m$^{-2}$ 
 (Figure~\ref{taylor_fig3} left, (Akahori \& Ryu 2010)), 
 while RM$_{\rm rms}$ through a number of  filaments up to z $\sim$ a few would reach several rad\,m$^{-2}$
  (Figure~\ref{taylor_fig3} middle, (Akahori \& Ryu 2011)). 
  Interestingly, such RM is comparable to the  estimate of extragalactic contribution to the  observed RM
  of  6 -- 15 rad\,m$^{-2}$ (Schnitzeler 2010, Hammond et al.\ 2012). 
  
 Simulations predict that the IGMF in filaments would induce RM with a flat second-order structure 
 function (SF) of $\sim$100 rad$^2$\,m$^{-4}$ for angular separation of r $\gtrsim$ 0$^{\circ}$.1
 (Figure~\ref{taylor_fig3} right). 
 Toward high galactic latitudes where the Galactic contribution is minimum, the Galactic magnetic field, on the 
 other hand, should produce substantially smaller and steeper SF in angular scale (Akahori et al.\ 2013). 
 Observed SFs toward  high latitude, while poorly sampled, are consistent with a flat SF on 
 the smallest scales (Mao et al.\ 2010, Stil et al.\ 2011), suggesting a  contribution from the IGMF on scales below 1$^{\circ}$.
 
 Directions toward high Galactic latitude and toward outside of galaxy clusters should be chosen for 
 deep-field surveys to detect the RM  of the IGMF in cosmic filaments. 
 The RM due to the Galactic magnetic field could be substantially reduced with spatial filtering to remove the large scale variance (Akahori et al.\ 2014). 
  Simulations have shown that a RM data set with a sky density of several 100 to 1000 per
 square degree and with RM precision of $\sim$1 rad~m$^{-2}$ is required to accurately reconstruction the structure function of RM
 variance due to the IGMF in the cosmic web (Akahori et al.\ 2014).  An SKA1-MID deep polarization survey to below 100 nJy with
 band 2 \& 3 would yield formal errors in RM of 1 rad~m$^{-2}$ for sources with polarized flux density greater than 1.3 $\mu$Jy. 
 At this polarized flux density the source density will approach the required 1000 per square degree (see Figure~\ref{taylor_fig1}). 
 A significant fraction of this population will be galaxies.  The alignment of polarization position angle with optical minor axis 
seen in local disk galaxies (Stil et al.\ 2009) offers the possibility to also use polarization and optical alignments of galaxies as another
powerful probe of  weak Faraday Rotation from large scale structure.

\section{Technical Summary}

The Cosmic Magnetism Deep Fields are intended to be observed commensally with the continuum survey deep fields 
using SKA1-MID over a frequency range of 950 - 3050 MHz (Band 2 \& 3).  
Band 3 will sample the regime where internal  Faraday depolarization
effects are negligible and alignment of polarization and geometric properties of the galaxy are preserved.   
Band 2 samples the depolarization regime and allows measurement of internal Faraday processes from galaxy structure 
(Heald et al.\ 2014). 
The significant depolarization in band 2 for nearby galaxies will be substantially reduced at z >1.   
Emission from galaxies in SKA1-MID Band 1 will suffer very significant depolarization for the red shifts available to SKA1.   
SKA1 Band 1 and 2  frequency coverage will  provide formal precision better than RM of 1 rad m$^{-2}$ down to polarized intensity 
of 1 $\mu$Jy\,beam$^{-1}$. 
Integration times sufficient for a sensitivity of 
75 nJy\,beam$^{-1}$ are required per field at a resolution of 1 arcsec. 
At this resolution the continuum (monochromatic)  rms noise for SKA1-MID is expected to be twice (three times) 
that of the naturally weighted visibility data for Band 2, given current array configuration projections. 

Star-forming galaxies with partly ordered regular fields are expected to have Faraday depth of  
$\sim$ 50 rad~m$^{-2}$ at $z\sim 0.5$. Large-scale field reversals may exist in those epochs and 
lead to several Faraday Depth components with different signs, so that a resolution of
50 rad m$^{-2}$ is required,  Band 2 \& 3 combine to provide Faraday depth resolution of 31 rad~m$^{-2}$

The majority of galaxies are expected of exhibit integrated polarized fraction of order 1 to a few percent, so linear  polarization purity of 0.1\% across 
the field of view is required.   Circular polarization purity of 0.01\% would allow circular 
polarization to be studied for the fist time in a vast number of AGN.

Imaging of an area of the order 10 square degrees is required to provide statistical sampling of magnetic fields in galaxies and AGN 
and for sampling of large-scale structure for detection of the cosmic magnetic web.
We expect a several tens of thousands of objects detected in polarization for study of the evolution of intrinsic magnetic fields with cosmic time. 
The dense grid of thousands of sources per square degree will probe the ``foreground'' cosmic magnetic web.

Both continuum and  spectral imaging will be required in full polarization for these data with a coarse spectral resolution of 
$\Delta\nu$ = 1 MHz.
Processing of the polarization deep fields will require some additions to a standard imaging pipeline. Channelized 
polarization products will need to be formed in full Stokes with high dynamic range per channel.
To avoid contamination from bright sources in the fields a dynamic range requirement of 10$^6$ (10$^5$) in total intensity 
(polarization) is required for continuum imaging and 10$^4$ (10$^3$) per channel for spectral data. 
To achieve this 
dynamic range, full Stokes A-projection will be required in addition to standard W-projection. Direction-dependent 
calibration will be necessary, followed by multifrequency synthesis.

These fully calibrated polarization data can then be transformed from frequency into Faraday dispersion cubes. Standard 
RM-synthesis would utilize the channelized QU image cube (Brentjens \& de Bruyn 2005).  However to
achieve full dynamic range  in the Faraday dispersion cube images from the broad bandwidth will require multi-frequency synthesis 
techniques to be adapted to Stokes Q, U data. Integration of multi-frequency synthesis and Faraday dispersion imaging would
require Faraday de-dispersion to operate on the visibilities prior to imaging and deconvolution.   
\makeatletter
\def\@biblabel#1{}
\makeatother\bibliographystyle{apj}

\end{document}